\documentclass[12pt]{article}

\usepackage{a41}

\usepackage{palatino}
\usepackage{xcolor}
\usepackage{amsmath}
\usepackage{booktabs}
\usepackage{microtype}
\usepackage{cite}
\usepackage{amsmath}
\usepackage{amssymb}
\usepackage{color}
\usepackage[rflt]{floatflt}
\usepackage{float}

\usepackage[rflt]{floatflt}
\usepackage{float}
\usepackage{slashed}
\def\b0{\beta_0}

\usepackage{array}

\usepackage[english]{babel}
\usepackage[latin1]{inputenc}
\usepackage[T1]{fontenc}
\usepackage{ae}

\usepackage{url}

\usepackage{amsmath, amsthm, amssymb}
\newtheorem{thm}{Theorem}[section]

\newtheorem{definition}[thm]{Definition}

\newcommand{\ep}{\varepsilon}

\usepackage{rotating}

\usepackage{graphicx}
\newcounter{mmacnt}
\def\restartmma{\setcounter{mmacnt}{0}}
\restartmma \catcode`|=\active
\def|#1|{\mathrm{#1}}
\catcode`|=12
\newenvironment{mma}{
 \par\smallskip
 \catcode`|=\active
 \parskip=0pt\parindent=0pt % locally
 \small
 \def\In##1\\{%
\def\linebreak{\hfill\break\null\qquad}%
\refstepcounter{mmacnt}
\hangindent=2.5em\hangafter=0
\leavevmode
\llap{\tiny\sffamily n[\arabic{mmacnt}]:=\kern.5em}%
\mathversion{bold}\footnotesize$\displaystyle##1$\normalsize
\mathversion{normal}\par
 }%
 \def\Print##1\\{%
\def\linebreak{\hfill\break}%
\hangindent=2.5em\hangafter=0
\leavevmode ##1\par}%
 \def\Out##1\\{%
\def\linebreak{$\hfill\break\null\hfill$}%
\kern\abovedisplayskip\par
\hangindent=2.5em\hangafter=0
\leavevmode
\llap{\tiny\sffamily Out[\arabic{mmacnt}]=\kern.5em}
\footnotesize$\displaystyle##1$\normalsize\hfill\null\par
\kern\belowdisplayskip
 }%
 \def\Warning##1##2\\{%
\def\linebreak{\hfill\break}%
\hangindent=2.5em\hangafter=0
\leavevmode
{\scriptsize##1 : ##2}\par}%
}{%
 \par\smallskip
}

\usepackage{color}

\newenvironment{fshaded}{%
\MakeFramed {\FrameRestore}
}%
{\endMakeFramed}

\def\b0{\beta_0}

\def\Gp0{{\Gamma^{'}_0}}
\def\Gp1{{\Gamma^{'}_1}}
\def\Gp2{{\Gamma^{'}_2}}

\allowdisplaybreaks[4]

\begin{document}
\setlength{\baselineskip}{0.515cm}

\sloppy
\thispagestyle{empty}
\begin{flushleft}
DESY 20--044
%%\hfill {\tt arXiv:2002.xxxxx[gr-qc]}
\\
DO--TH 20/02\\
SAGEX--20--06\\
\end{flushleft}

\mbox{}
\vspace*{\fill}
\begin{center}

{\Large\bf Testing binary dynamics in gravity}

\vspace*{3mm}
{\Large\bf  at the sixth post-Newtonian level}

\vspace{3cm} \large
{\large J.~Bl\"umlein$^a$, A.~Maier$^a$, P.~Marquard$^a$, and G.~Sch\"afer$^b$}

\vspace{1.cm}
\normalsize
{\it $^a$Deutsches Elektronen--Synchrotron, DESY,}\\
{\it   Platanenallee 6, D--15738 Zeuthen, Germany}

\vspace*{2mm}
{\it $^b$Theoretisch-Physikalisches Institut, Friedrich-Schiller-Universit\"at, \\
Max-Wien-Platz 1, D--07743 Jena, Germany}\\

%%\today

\end{center}
\normalsize
\vspace{\fill}
\begin{abstract}
\noindent
We calculate the motion of binary mass systems in gravity up to the sixth post--Newtonian order to the $G_N^3$ terms
ab initio using momentum expansions within an effective field theory approach based on Feynman amplitudes in harmonic 
coordinates. For these contributions we construct a canonical transformation to isotropic and to EOB coordinates at 5PN 
and agree with the results in the literature \cite{Bern:2019nnu,Damour:2019lcq}. At 6PN we compare to the Hamiltonians 
in isotropic coordinates either given in \cite{Bern:2019nnu} or resulting from the scattering angle. We find a canonical 
transformation from our Hamiltonian in harmonic coordinates to \cite{Bern:2019nnu}, but not to \cite{Damour:2019lcq}. 
This implies that we also agree on all observables with \cite{Bern:2019nnu} to the sixth post--Newtonian order to $G_N^3$.
\end{abstract}

\vspace*{\fill}
\noindent
% \numberwithin{equation}{section}
%%%%%%%%%%%%%%%%%%%%%%%%%%%%%%%%%%%%%%%%%%%%%%%%%%%%%%%%%%%%%%%%%%%%%%%%%%%%%%%%%%%%%%%%%%%%%%%%%%%%%%%%%%%%%%%%%%%%%%%%%%%%%%%%%%%
\newpage
%----------------------------------------------------------------------------------------------------------------
\section{Introduction}
\label{sec:1}
%----------------------------------------------------------------------------------------------------------------

\vspace*{1mm}
\noindent
The observation of gravitational wave signals coming from merging black holes and neutron stars \cite{LIGO} 
is a milestone in astrophysics. The detectors reach higher and higher sensitivity \cite{PROJECT} and do thus 
yield precision data to be confronted with precision calculations in general relativity. The process of 
merging of massive gravitating objects can be analytically described by the post--Newtonian (PN) approximation
in the region of lower velocities. Currently the state of the art is the fourth post--Newtonian approximation
\cite{FOURTH,Foffa:2016rgu,Blumlein:2020pog} with first results at the fifth post--Newtonian level 
\cite{Foffa:2019hrb,Blumlein:2019zku,Bini:2019nra}.

For scattering processes, i.e. at high energies of the massive objects, the post--Minkowskian (PM) approximation 
\cite{PM1,Damour:2017zjx,Bern:2019nnu,Bern:2019crd,Damour:2019lcq} holds, where the third 
post--Minkowskian level \cite{Bern:2019nnu,Bern:2019crd} has been reached recently. It is possible to derive all 
contributions to the given post--Newtonian level for a given power in the Newton's gravitation constant $G_N$ 
and to obtain in this way cross checks between different calculation methods.

The method of amplitudes provides a powerful approach to many scattering processes in the Standard Model of 
elementary particles, and more recently also to classical gravity, cf.~e.g.~\cite{Bern:2019nnu,Bern:2019crd}. 
Any of the different approaches to calculate the dynamics of two--body systems has to be constantly tested with 
respect to their validity in view of other approaches since the goal is the consistent derivation of predictions for 
observables at higher and higher post--Newtonian and post--Minkowskian levels. 

In course of this, Ref.~\cite{Damour:2019lcq} conjectured a modified 3PM result having a softer high-energy behavior 
differing of the results of \cite{Bern:2019crd,Bern:2019nnu} in the contribution to the scattering angle $\chi$ at the 
level of the sixth post--Newtonian level, which can be obtained by a momentum expansion of Hamiltonians calculated to 
the third post--Minkowskian ($O(G_N^3/r^3)$) level, and proposed to test this by calculations ab initio.

It is known that momentum expansions of the Hamiltonian describing the two--body system allow the reconstruction of the 
post--Minkowskian Hamiltonians in finite terms \cite{Blumlein:2019bqq}. The momentum expansions of post--Minkowskian 
Hamiltonians have to agree with the respective contributions to the post--Newtonian Hamiltonians, which can be calculated 
using effective field theory methods based on Feynman amplitudes. 

In this note we will extend earlier work up to the level of 4PN \cite{Blumlein:2020pog} to the sixth post--Newtonian 
order to  $O(G_N^3/r^3)$ to answer the question raised in Ref.~\cite{Damour:2019lcq}  using the Hamiltonian formalism, 
cf.~\cite{Schafer:2018kuf}. In Section~\ref{sec:2} we describe the calculation of the corresponding contributions to the 
Hamiltonian at 6PN in harmonic coordinates. Different Hamiltonians based on harmonic, isotropic and EOB coordinates to 5PN 
and $O(G_N^3)$ are related to each other by construction of the canonical transformations between them in 
Section~\ref{sec:3}. 
Here we find mutual agreements. The comparison at 6PN to $O(G_N^3)$ is performed in Section~\ref{sec:4} and 
Section~\ref{sec:5} contains the conclusions.
%----------------------------------------------------------------------------------------------------------------
\section{Details of the calculation}
\label{sec:2}
%----------------------------------------------------------------------------------------------------------------

\vspace*{1mm}
\noindent
Starting from the Einstein--Hilbert Lagrangian, we parameterize the metric $g_{\mu\nu}$ following 
Ref.~\cite{Kol:2007bc}\footnote{Following the ideas in \cite{Goldberger:2004jt}.} 
in terms of scalar, vector and tensor fields. We work in the harmonic gauge. The path-integral representation yields the 
Feynman rules and the diagrams are generated using {\tt QGRAF} \cite{Nogueira:1991ex}. We work thoroughly in $D = 4 - 2\ep$
dimensions and calculate all contributions needed 
to 6PN up to the order $G_N^3/r^3$ in the effective field theory approach. Here $G_N$ denotes Newton's constant. The 
Lorentz algebra is carried out using {\tt Form} \cite{FORM} and we perform the integration by parts (IBP) reduction to 
master integrals using the code {\tt Crusher} \cite{CRUSHER}. 
Many of the technical details used have been described in detail in Refs.~\cite{Blumlein:2019zku,Blumlein:2020pog}.

%--------------------------------------------------------------------------------------------------------
\begin{table}[H]\centering
\begin{tabular}{rrrrrrr}
\toprule
\multicolumn{1}{c}{\#loops }            &
\multicolumn{1}{c}{\tt{QGRAF}}        &
\multicolumn{1}{c}{source irred}           &
\multicolumn{1}{c}{no source loops}  &
\multicolumn{1}{c}{no tadpoles}  &
\multicolumn{1}{c}{masters} &
\multicolumn{1}{c}{CPU time [min]} 
\\
\midrule
  0  &      3  &      3 &      3 &      3 & 0 &     9   \\
  1  &     72  &     72 &     72 &     72 & 1 &   163   \\
  2  &   4322  &   4322 &   4322 &   3512 & 1 & 41796   \\
\bottomrule
\label{TAB1}
\end{tabular}
\caption[]{\sf Numbers of contributing diagrams at the different loop levels and master integrals and 
the CPU times needed for the calculation.} 
\label{TAB1}
\end{table}
%--------------------------------------------------------------------------------------------------------

In Table~\ref{TAB1} we summarize the complexity of the present calculation, with contributions up to 
two loops. From the graphs generated by {\tt QGRAF} one has to remove the source irreducible graphs,
graphs with source loops and tadpoles.\footnote{To be consistent with previous papers 
\cite{Blumlein:2019zku,Blumlein:2020pog} we give the columns 2--4 individually, despite the numbers do not change.}
In this way the 4397 initial diagrams reduce to 3587 diagrams.
The computation time amounts to about 30 days, including the time for the IBP reduction, on an {\tt Intel(R) Xeon(R) CPU 
E5-2643 v4} and it grows exponentially with the loop order. Most of the CPU time is needed to perform the time derivatives.
Only one master integral contributes, see \cite{Foffa:2016rgu,Blumlein:2019zku}.

In the present calculation no tail--terms contribute, since they emerge only from $G_N^4/r^4$ on \cite{Blanchet:1987wq,
Bini:2017wfr,Foffa:2019eeb}.

One first obtains a Lagrange function of $m$th order still  containing the accelerations $a_i$ and time derivatives 
thereof. They are removed using double zero insertions \cite{Damour:1985mt} and the linear accelerations by a shift 
\cite{DW,Damour:1985mt,Damour:1990jh}, cf.~\cite{Blumlein:2020pog}. A Legendre transformation leads then to the Hamiltonian.
%----------------------------------------------------------------------------------------------------------------
\section{The Hamiltonians to 5PN and \boldmath $O(G_N^3)$}
\label{sec:3}
%----------------------------------------------------------------------------------------------------------------

\vspace*{1mm}
\noindent
Let us first relate the Hamiltonian to 5PN in harmonic coordinates 
$\hat{H}_{\rm harm}^{\rm 5PN}$ to the ones from \cite{Bern:2019crd} $\hat{H}_{\rm isotr}^{\rm 5PN}$ and the EOB
Hamiltonian \cite{Bini:2019nra}. The Hamiltonian of \cite{Bern:2019crd}, expanded to the sixth 
post--Newtonian order, is given by
%--------------------------------------------------------------------------------------------------------------------------
\begin{eqnarray}
\hat{H}^{\rm B} &=& \hat{H}_{\rm N} + \sum_{k=1}^6 \hat{H}_{\rm kPN}^{\rm B} + O({\rm 7PN})
\\
%---------------
\label{eq:H6PNexpI}
\hat{H}_{\rm N} &=& \nu \left(\frac{1}{2} p^2 - u \right)
\\
%---------------
\hat{H}_{\rm 1PN}^{\rm B} &=& \nu \Biggl(
\frac{1}{8} (-1+3 \nu ) p^4
+\frac{1}{2} (-3-2 \nu ) p^2 u
+\frac{1}{2} (1+\nu ) u^2 \Biggr)
\\
%---------------
\hat{H}_{\rm 2PN}^{\rm B} &=& \nu \Biggl(
\frac{1}{16} \big(
        1-5 \nu +5 \nu ^2\big) p^6
+\frac{1}{8} \big(
        5-20 \nu -8 \nu ^2\big) p^4 u
+\frac{1}{4} \big(
        10+27 \nu +3 \nu ^2\big) p^2 u^2
\nonumber\\ 
&& -\frac{1}{4} (1+6 \nu ) u^3 \Biggr)
\\
%---------------
\hat{H}_{\rm 3PN}^{\rm B} &=& 
\nu \Biggl[\Biggl(
        -\frac{5}{128}+\frac{35 \nu }{128}-\frac{35 \nu ^2}{64}+\frac{35 \nu ^3}{128}\Biggr) p^8
+\Biggl(
        -\frac{7}{16}+\frac{21 \nu }{8}-3 \nu ^2-\nu ^3\Biggr) p^6 u
\nonumber\\ &&
+\Biggl(
        -\frac{27}{16}+\frac{147 \nu }{16}
+\frac{111 \nu ^2}{8}+\frac{15 \nu ^3}{16}\Biggr) p^4 u^2
+\frac{1}{8} \Biggl(
        -25-130 \nu -107 \nu ^2\Biggr) p^2 u^3
\\ 
\hat{H}_{\rm 4PN}^{\rm B} &=& 
\Biggl(
        \frac{7}{256}-\frac{63 \nu }{256}+\frac{189 \nu ^2}{256}-\frac{105 \nu ^3}{128}+\frac{63 \nu ^4}{256}\Biggr) p^{10}
+\Biggl(
        \frac{45}{128}-\frac{45 \nu }{16}+\frac{51 \nu ^2}{8}-3 \nu ^3-\nu ^4\Biggr) p^8 u
\nonumber\\ &&
+\Biggl(
        \frac{13}{8}-\frac{379 \nu }{32}+\frac{453 \nu ^2}{32}+\frac{337 \nu ^3}{16}+\frac{35 \nu ^4}{32}\Biggr) p^6 u^2
+\Biggl(
        \frac{105}{32}-\frac{4049 \nu }{160}-\frac{2589 \nu ^2}{32}
\nonumber\\ &&
-\frac{487 \nu ^3}{16}\Biggr) p^4 u^3 \Biggr]
\\ 
\hat{H}_{\rm 5PN}^{\rm B} &=&
\nu \Biggl[\Biggl(
        -\frac{21}{1024}+\frac{231 \nu }{1024}-\frac{231 \nu ^2}{256}+\frac{1617 \nu ^3}{1024}-\frac{1155 \nu ^4}{1024}
+\frac{231 \nu ^5}{1024}\Biggr) p^{12}
+\Biggl(
        -\frac{77}{256}
\nonumber\\ &&
+\frac{385 \nu }{128}-10 \nu ^2
+\frac{95 \nu ^3}{8}-\frac{5 \nu ^4}{2}-\nu ^5\Biggr) p^{10} u
+\Biggl(
        -\frac{425}{256}+\frac{3915 \nu }{256}-\frac{2475 \nu ^2}{64}+\frac{2625 \nu ^3}{256}
\nonumber\\ && +\frac{7107 \nu 
^4}{256}
+\frac{315 \nu ^5}{256}\Biggr) 
p^8 u^2
+\Biggl(
        -\frac{273}{64}+\frac{178553 \nu }{4480}-\frac{30993 \nu ^2}{640}-\frac{1527 \nu ^3}{8}
\nonumber\\  &&
-\frac{6607 \nu 
^4}{128}\Biggr) 
p^6 u^3 \Biggr]
\\
\hat{H}_{\rm 6PN}^{\rm B} 
&=& 
\nu \Biggl[
\Biggl(
        \frac{33}{2048}-\frac{429 \nu }{2048}+\frac{2145 \nu ^2}{2048}-\frac{1287 \nu ^3}{512}+\frac{3003 \nu ^4}{1024} 
-\frac{3003 \nu ^5}{2048}+\frac{429 \nu ^6}{2048}\Biggr\} p^{14} \nonumber\\ &&
+ \Biggl(
        \frac{273}{1024}-\frac{819 \nu }{256}+\frac{1785 \nu ^2}{128}
-\frac{105 \nu ^3}{4}+\frac{75 \nu ^4}{4}-\frac{3 \nu ^5}{2}-\nu ^6\Biggr) p^{12} u
   + \Biggl(    \frac{441}{256}-\frac{9849 \nu }{512}
\nonumber\\ &&
+\frac{36129 \nu ^2}{512}
-\frac{10821 \nu ^3}{128}-\frac{2505 \nu ^4}{256}+\frac{17175 \nu ^5}{512}+\frac{693 \nu ^6}{512}\Biggr) p^{10} u^2 
\Biggr]
+ f_6(\nu) p^8 u^3,
\label{eq:H6PNexpF}
\end{eqnarray}
%--------------------------------------------------------------------------------------------------------
with
%--------------------------------------------------------------------------------------------------------
\begin{eqnarray}
\label{eq:F6}
f_6(\nu) &=& \sum_{k=0}^5 f_{6,k} \nu^k.
\end{eqnarray}
%--------------------------------------------------------------------------------------------------------
with
%--------------------------------------------------------------------------------------------------------
\begin{eqnarray}
f_6(\nu) &\equiv& 
f_6^{\rm B}(\nu) = 
        \frac{2805 \nu}{512} -\frac{1947527 \nu^2 }{32256}
+\frac{3093791 \nu ^3}{17920}+\frac{5787 \nu ^4}{320}
-\frac{168131 \nu ^5}{512}-\frac{19425 \nu ^6}{256}.
\end{eqnarray}
%--------------------------------------------------------------------------------------------------------
Here we normalized 
%--------------------------------------------------------------------------------------------------------
\begin{eqnarray}
\hat{H} = \frac{H}{M},~~~M = m_1 + m_2,~~~p = \frac{P}{\mu},~~~\mu = \frac{m_1 m_2}{m_1+m_2},~~~u = \frac{G_N M}{r},~~~ 
\nu = \frac{\mu}{M}
\end{eqnarray}
%--------------------------------------------------------------------------------------------------------
and introduced a series of variables for dimensionless representations in the following and set the velocity of light $c=1$. 
If not stated otherwise
we will use $u \equiv 1/r$ synonymously in the following.

In the Schwarzschild approximation the Hamiltonian in isotropic coordinates reads \cite{SCHMUTZER1,WEINB}
%--------------------------------------------------------------------------------------------------------
\begin{eqnarray}
\hat{H} = \nu \frac{1-u/2}{1+u/2} \sqrt{1+ \frac{p^2}{(1+u/2)^4}}
\end{eqnarray}
%--------------------------------------------------------------------------------------------------------
providing a check on all contributions of $O(\nu)$ in (\ref{eq:H6PNexpI}--\ref{eq:H6PNexpF}).

The general structure of the generators of the canonical transformation is given by
%--------------------------------------------------------------------------------------------------------
\begin{eqnarray}
g(p^2, p.r, u, \nu) &=& p.r \Biggl\{ \sum_{k=0}^{N_1} u^k 
\Biggl[          \sum_{l=0}^{N_2-k} \alpha_{k,l,1}(\nu) (p^2)^{N_2-l} (p.n)^{2l}  
+  \frac{1}{\ep} \sum_{l=0}^{N_2-k} \alpha_{k,l,2}(\nu) (p^2)^{N_2-l} (p.n)^{2l} 
\nonumber\\ && 
+  \ln\left(\frac{r}{r_0}\right) \sum_{l=0}^{N_2-k} \alpha_{k,l,3}(\nu) (p^2)^{N_2-l} (p.n)^{2l} 
\Biggr] \Biggr\},
\label{eq:gstr}
\end{eqnarray}
%--------------------------------------------------------------------------------------------------------
with $p.n = p.r/r$ and
%--------------------------------------------------------------------------------------------------------
\begin{eqnarray}
r_0 = \frac{e^{-\gamma_E/2}}{2 \sqrt{\pi} \mu_1},
\end{eqnarray}
%--------------------------------------------------------------------------------------------------------
where $\gamma_E$ is the Euler--Mascheroni constant and $\mu_1$ a mass scale appearing in $D$-dimensional 
regularization. We allow also for pole- and logarithmic terms, which arise e.g. in harmonic coordinates.
In the present calculation we have $N_1 = 3$ and $N_2 = 7$.

We use the method of Lie--series \cite{GROEBNER,MITTELSTAEDT} to determine the generators of the canonical transformation, 
see Sections~2.5--4 of Ref.~\cite{Blumlein:2020pog}.
%--------------------------------------------------------------------------------------------------------
\begin{eqnarray}
\label{eq:HTR}
H_{\rm isotr}^{\rm \leq G^3_N}(y) = \exp[D_g] H_{\rm harm}^{\rm \leq G^3_N}(y) = H_{\rm harm}^{\rm \leq G^3_N}(y'),
\end{eqnarray}
%--------------------------------------------------------------------------------------------------------
where $y$ and $y'$ denote the respective canonical coordinates and the differential operator $D_g$ is defined in Eq.~(34) 
of \cite{Blumlein:2020pog}.

The generators of this transformation are given to the fifth post--Newtonian order by\footnote{Note that the 
results
in \cite{Bern:2019crd} deliver all post--Newtonian terms to 2PN and for higher post--Newtonian orders they are
supposed to yield all contributions up to $O(G^3_N)$. Comparing to our 4PN complete result \cite{Blumlein:2020pog} we have 
confirmed this to 4PN already.} 
%--------------------------------------------------------------------------------------------------------
\begin{eqnarray}
g_{1} &=& 
\frac{p.r}{2 r} \nu
\\
%-------------------------------------
g_{2} &=& \frac{p.r}{4r} \Biggl\{
-\frac{1}{2} \nu ^2 \left(p^2 - (p.n)^2 \right)
+\frac{1}{r} \left(1-3 \nu +\nu ^2\right)
\Biggr\}
\\
g_{3} &=& {\frac{p.r}{r}} \Biggl\{
-\frac{1}{48} \nu ^3 \left(3 p^4 + p^2 (p.n)^2 +3 (p.n)^4 \right)
+\frac{1}{r} \Biggl[
        \left(
                \frac{9 \nu }{4}-\frac{5 \nu ^2}{8}+\frac{\nu ^3}{16}\right) p^2
\nonumber\\ &&
        +\left(
                -\frac{7 \nu }{12}-\frac{11 \nu ^2}{2}+\frac{5 \nu ^3}{48}\right) (p.n)^2
\Biggr]
+\frac{1}{r^2} \left(
        \frac{2789 \nu }{144}
        +\frac{5 \nu ^2}{16}
        +\frac{\nu ^3}{16}
        -\frac{7 \nu  \pi ^2}{8}
\right) - {\frac{17 \nu}{r^2} L}
\Biggr\}
\\
%-------------------------------------
g_{4} &=& \frac{p.r}{r} \Biggl\{ 
-\frac{1}{128} \nu ^3 \left[(-96+5 \nu ) p^6 +\frac{5}{3} (4+\nu ) p^4 (p.n)^2 + (4+\nu ) p^2 (p.n)^4 - {5 \nu (p.n)^6}
\right]
\nonumber\\ &&
+\frac{1}{r} \Biggl[
        \left(
                \frac{309 \nu }{64}-\frac{635 \nu ^2}{64}-\frac{35 \nu ^3}{4}+\frac{\nu ^4}{32}\right) p^4
        +\left(
                -\frac{79 \nu }{96}+\frac{413 \nu ^2}{32}+\frac{797 \nu ^3}{48}+\frac{\nu ^4}{48}\right) p^2 (p.n)^2
\nonumber\\ &&
        +\left(
                -\frac{487 \nu }{960}-\frac{181 \nu ^2}{64}-\frac{142 \nu ^3}{15}+\frac{7 \nu ^4}{96}\right) (p.n)^4
\Biggr]
+\frac{1}{r^2} \Biggl[
         \Biggl(
                -\frac{824117 \nu }{14400}
                +\frac{341089 \nu ^2}{14400}
                -\frac{3 \nu ^3}{64}
\nonumber\\ &&
                +\frac{\nu ^4}{16}
                +\nu \left(\frac{643}{1024}
                -\frac{133}{64} \nu \right) \pi ^2
                +\frac{1}{15} \nu  (585+4 \nu ) L
        \Biggr) p^2
        + \Biggl(
                \frac{34973 \nu }{960}
                +\frac{151089 \nu ^2}{1600}
                +\frac{239 \nu ^3}{192}
\nonumber\\ && 
                +\frac{\nu ^4}{32}
                -\nu \left(\frac{879 }{1024}
                {+} \frac{69}{128} \nu\right) \pi ^2
                -2 \nu  (12+37 \nu ) L
        \Biggr) (p.n)^2
\Biggr]
\Biggr\}
\\
%-------------------------------------
g_{5} &=& \frac{p.r}{r} \Biggl\{
\Biggl(
        \frac{91 \nu ^3}{128}-\frac{97 \nu ^4}{64}-\frac{7 \nu ^5}{256}\Biggr) p^8
+\Biggl(
        -\frac{51 \nu ^3}{256}+\frac{337 \nu ^4}{768}-\frac{7 \nu ^5}{768}\Biggr) p^6 (p.n)^2
\nonumber\\ &&
+\Biggl(
        \frac{3 \nu ^3}{256}+\frac{17 \nu ^4}{256}-\frac{7 \nu ^5}{1280}\Biggr) p^4 (p.n)^4
+\Biggl(
        \frac{5 \nu ^3}{256}-\frac{25 \nu ^4}{256}-\frac{\nu ^5}{256}\Biggr) p^2 (p.n)^6
+\frac{7 \nu ^5}{256} (p.n)^8
\nonumber\\ &&
+\frac{1}{r} \Biggl[
        \Biggl(
                \frac{1689 \nu }{128}-\frac{5601 \nu ^2}{128}+\frac{475 \nu ^3}{16}+\frac{3543 \nu ^4}{128}+\frac{5 \nu 
^5}{256}\Biggr) p^6
        +\Biggl(
                -\frac{715 \nu }{192}+\frac{31 \nu ^2}{6}+\frac{439 \nu ^3}{48}
\nonumber\\ &&
-\frac{9001 \nu ^4}{96}+\frac{7 \nu 
^5}{768}\Biggr) p^4 (p.n)^2
        +\Biggl(
                \frac{19 \nu }{12}+\frac{3949 \nu ^2}{960}-\frac{8711 \nu ^3}{960}+\frac{116237 \nu ^4}{960}+\frac{49 \nu 
^5}{3840}\Biggr) p^2 (p.n)^4
\nonumber\\ &&      
  +\Biggl(
                -\frac{159 \nu }{224}-\frac{751 \nu ^2}{224}+\frac{2263 \nu ^3}{280}-\frac{89981 \nu ^4}{2240}+\frac{15 \nu 
^5}{256}\Biggr) (p.n)^6
\Biggr]
+\frac{1}{r^2} \Biggl[
        \Biggl(
                \frac{12576721 \nu }{705600}
\nonumber\\ &&        
        +\frac{912076073 \nu ^2}{2822400}
                +\frac{281619239 \nu ^3}{8467200}
                +\frac{41 \nu ^4}{4}
                +\frac{15 \nu ^5}{256}
                +\Biggl(
                        -\frac{5089 \nu }{4096}+\frac{15153 \nu ^2}{4096}-\frac{15387 \nu ^3}{1024}\Biggr) \pi ^2
\nonumber\\ &&                
+\Biggl(
                        \frac{2277 \nu }{20}-\frac{69887 \nu ^2}{210}-\frac{10001 \nu ^3}{120}\Biggr) L
        \Biggr) p^4
        + \Biggl(
                \frac{15911 \nu }{192}
                -\frac{775711 \nu ^2}{188160}
                +\frac{5196367 \nu ^3}{12544}
\nonumber\\ &&                
+\frac{5913 \nu ^4}{640}
                +\frac{47 \nu ^5}{1280}
                +\Biggl(
                        -\frac{6015 \nu }{4096}-\frac{2025 \nu ^2}{4096}+\frac{9375 \nu ^3}{1024}\Biggr) \pi ^2
                -\frac{1}{2} \nu  \big(
                        132-26 \nu +413 \nu ^2\big) L
        \Biggr) 
\nonumber\\ && \times
{(p.n)^4}
        + \Biggl(
                -\frac{12459777 \nu }{78400}
                -\frac{25111447 \nu ^2}{470400}
                -\frac{6705133 \nu ^3}{14700}
                -\frac{3875 \nu ^4}{192}
                +\frac{\nu ^5}{64}
       +\Biggl(
                        \frac{1269 \nu }{512}
\nonumber\\ &&
-\frac{1455 \nu ^2}{256}+\frac{8109 \nu ^3}{512}\Biggr) \pi ^2
                + \Biggl(
                        \frac{96 \nu }{5}+168 \nu ^2+\frac{43567 \nu ^3}{140}\Biggr) L
\Biggr) p^2 (p.n)^2
\Biggr]
\Biggr\}
\end{eqnarray}
%---------------------------------------------------------------------------------------------
with 
%--------------------------------------------------------------------------------------------------------
\begin{eqnarray}
L  \equiv L(r,r_0,\ep) = \ln\left(\frac{r}{r_0}\right) + \frac{1}{6 \ep}.
\end{eqnarray}
%--------------------------------------------------------------------------------------------------------
According to (\ref{eq:HTR}) the generators $g_i$ appear in the corresponding (multiple) Poisson brackets, cf.~e.g. 
\cite{Blumlein:2020pog}.

Likewise, we obtain a canonical transformation from harmonic coordinates to EOB coordinates given by the generators
$G_1$ to $G_4$ given in Ref.~\cite{Blumlein:2020pog} and
%--------------------------------------------------------------------------------------------------------
\begin{eqnarray}
\label{eq:g5EOB}
G_5^{\rm harm-EOB} &=&
p.r \nu \Biggl\{
        \frac{1}{768} \big(
                21-63 \nu +40 \nu ^2\big) p^{10}
%---     
   + \Biggl[
                \Biggl(
                \frac{55}{256}-\frac{941 \nu }{1536}+\frac{569 \nu ^2}
                     {512}-\frac{1927 \nu ^3}{768}+\frac{\nu ^4}{768}\Bigg) p^8
\nonumber\\ &&
                +\Biggl(
                        \frac{5}{32}-\frac{5 \nu }{64}-\frac{205 \nu ^2}{384}+\frac{145 \nu ^3}{768}+\frac{\nu 
^4}{4608}\Biggr) p^6 (p.n)^2
                +\Biggl(
                        -\frac{\nu }{4}+\frac{65 \nu ^2}{256}+\frac{29 \nu ^3}{128}
\nonumber\\ &&
+\frac{37 \nu ^4}{7680}\Biggr) p^4 (p.n)^4 
                +\Biggl(
                        \frac{5 \nu ^2}{256}-\frac{203 \nu ^3}{768}-\frac{7 \nu ^4}{1536}\Biggr) p^2 (p.n)^6
                +\frac{7 \nu ^3}{48} (p.n)^8
        \Biggr] \frac{1}{r}
%----
+ \Biggl[\Biggl(
                        \frac{443}{32}
\nonumber\\ &&
-\frac{9779 \nu }{192}+\frac{16783 \nu ^2}{384}+\frac{14675 \nu ^3}{384}-\frac{\nu 
^4}{384}\Biggr) p^6
                +\Biggl(
                        -\frac{269}{96}+\frac{87 \nu }{32}+\frac{6157 \nu ^2}{384}
\nonumber\\ &&
-\frac{100025 \nu ^3}{1152}+\frac{13 \nu 
^4}{576}\Biggr) p^4 (p.n)^2
                +\Biggl(
                        \frac{85}{48}+\frac{321 \nu }{160}-\frac{793 \nu ^2}{20}+\frac{484729 \nu ^3}{5760}-\frac{199 \nu 
^4}{1440}\Biggr) 
\nonumber\\ && \times
p^2 (p.n)^4
                +\Biggl(
                        -\frac{21}{32}-\frac{5309 \nu }{2240}+\frac{21565 \nu ^2}{1344}-\frac{63677 \nu ^3}{2688}+\frac{13 
\nu ^4}{128}\Biggr) (p.n)^6
        \Biggr] \frac{1}{r^2}
%---
\nonumber\\ && 
+ \Biggl[
                \Biggl(
                        \frac{44527097}{1411200}
                        +\frac{167668187 \nu }{470400}
                        -\frac{52415933 \nu ^2}{1693440}
                        -\frac{7121 \nu ^3}{640}
                        -\frac{77 \nu ^4}{1280}
\nonumber\\ &&                        
+\frac{\big(
                                -5089+13448 \nu -55388 \nu ^2\big)}{4096} \pi^2
                        +\frac{\big(
                                191268-606451 \nu -142135 \nu ^2\big)}{1680} L
                \Biggr) p^4
\nonumber\\ &&                
+ \Biggl(
                        -\frac{4916747}{29400}
                        +\frac{317945611 \nu }{2822400}
                        -\frac{279857647 \nu ^2}{705600}
                        +\frac{105479 \nu ^3}{5760}
                        +\frac{23 \nu ^4}{720}
\nonumber\\ &&                    
    +\frac{3 \big(
                                3384-10235 \nu +25696 \nu ^2\big) }{4096} \pi^2
                        +\frac{1}{140} \big(
                                2688+12705 \nu +40921 \nu ^2\big) L
                \Biggr) p^2 (p.n)^2
\nonumber\\ &&
 + \Biggl(
                        \frac{1136021}{13440}
                        -\frac{73782167 \nu }{564480}
                        +\frac{67938979 \nu ^2}{564480}
                        -\frac{7073 \nu ^3}{1440}
                        +\frac{697 \nu ^4}{11520}
\nonumber\\ &&                        
+\frac{5 \big(
                                -1203+1353 \nu +7708 \nu ^2\big)}{4096} \pi^2
                        +\frac{1}{4} \big(
                                -264+292 \nu -171 \nu ^2\big) L
                \Biggr) (p.n)^4
        \Biggr] \frac{1}{r^3}
\Biggr\}
\nonumber\\ && 
+ O\left(\frac{1}{r^4}\right).
\end{eqnarray}
%--------------------------------------------------------------------------------------------------------
Since the canonical transformations between all the three Hamiltonians exist, they are physically equivalent to 
the level of 5PN and $O(G_N^3)$ and agree in the respective contributions for {\it all} observables, including 
the scattering angle.
%----------------------------------------------------------------------------------------------------------------
\section{The comparison at 6PN to \boldmath $O(G_N^3)$}
\label{sec:4}
%----------------------------------------------------------------------------------------------------------------

\vspace*{1mm}
\noindent
It has been shown in \cite{Damour:2017zjx} that for conservative dynamics to 3PM one can relate the different 
post--Minkowskian contributions for the scattering angle to the Hamiltonian, see also~\cite{Antonelli:2019ytb}. 
We will apply this to the case of isotropic coordinates.

The scattering angle is given by the following post--Minkowskian series
%--------------------------------------------------------------------------------------------------------
\begin{eqnarray}
\frac{1}{2}\chi  = \sum_{k=1}^\infty \frac{1}{j^k} \chi_k,
\end{eqnarray}
%--------------------------------------------------------------------------------------------------------
where 
%--------------------------------------------------------------------------------------------------------
\begin{eqnarray}
j = \frac{J}{G_N m_1 m_2}
\end{eqnarray}
%--------------------------------------------------------------------------------------------------------
denotes the normalized angular momentum. We apply the notation in \cite{Antonelli:2019ytb} and \cite{Damour:2019lcq}.
For the contribution $\chi_3$ the following contributions have been obtained to this order:
%--------------------------------------------------------------------------------------------------------
\begin{eqnarray}
\chi_3(\gamma,\nu)  &=& \chi_3^{\rm Schw}(\gamma,\nu) - \frac{2 \nu p_\infty}{\Gamma^2} \overline{C}(\gamma),
\end{eqnarray}
%--------------------------------------------------------------------------------------------------------
with \cite{Damour:2017zjx}
%--------------------------------------------------------------------------------------------------------
\begin{eqnarray}
\chi_3^{\rm Schw}  &=& \frac{1}{3 p_\infty^3} [64 p_\infty^6 +72 p_\infty^4 +12 p_\infty^2 - 1]
\end{eqnarray}
%--------------------------------------------------------------------------------------------------------
and \cite{Damour:2019lcq}, Eqs.~(3.72, 3.73),
%--------------------------------------------------------------------------------------------------------
\begin{eqnarray}
\overline{C}^{\rm 5PN}(\gamma) &=& 4 + 18 p^2_\infty + \frac{91}{10} p^4_\infty - \frac{69}{140} p_\infty^6,
\\ 
\overline{C}^{\rm 6PN}(\gamma) &=&  \overline{C}^{\rm 5PN}(\gamma) + \overline{C}^{\rm 6PN, B(D)}(\gamma)
\end{eqnarray} 
%--------------------------------------------------------------------------------------------------------
and
%--------------------------------------------------------------------------------------------------------
\begin{eqnarray}
\label{eq:CB}
\overline{C}^{\rm 6PN, B}(\gamma) &=& -\frac{1447}{10080} p_\infty^8,
\\
\label{eq:CD}
\overline{C}^{\rm 6PN, D}(\gamma) &=& -\frac{233}{672} p_\infty^8.
\end{eqnarray}
%-------------------------------------------------------------------------------------------------------- 
One has
%--------------------------------------------------------------------------------------------------------
\begin{eqnarray}
p_\infty = \sqrt{\gamma^2 - 1},~~~~~~~\Gamma = \sqrt{1 + 2\nu (\gamma-1)},
\end{eqnarray}
%-------------------------------------------------------------------------------------------------------- 
where
%--------------------------------------------------------------------------------------------------------
\begin{eqnarray}
\gamma = \frac{1}{m_1 m_2} \left[ E_1 E_2 + p^2 \right],
\end{eqnarray}
%-------------------------------------------------------------------------------------------------------- 
with $E_{1,2} = \sqrt{m_{1,2}^2 +p^2}$. With these relations one can perform a momentum expansion in $p$ of the 
contribution $\chi_3$ to the scattering angle to 6PN.

Using the functions (3.69, 3.71) in \cite{Damour:2017zjx} one obtains the difference term to the Hamiltonian
%--------------------------------------------------------------------------------------------------------
\begin{eqnarray}
\hat{H}_{\rm isotr}^{\rm 6PN, D-B} = -\frac{64}{315} \nu^2 p^8 u^3
\end{eqnarray}
%-------------------------------------------------------------------------------------------------------- 
contributing at the sixth post--Newtonian order. It has to be added to (\ref{eq:H6PNexpF})
and yields
%-------------------------------------------------------------------------------------------------------- 
\begin{eqnarray} 
f_6^D(\nu) = f_6^B(\nu) -\frac{64}{315} \nu^2,
\label{eq:f6D}
\end{eqnarray} 
%--------------------------------------------------------------------------------------------------------
cf.~(\ref{eq:F6}). The numerical coefficient of $\hat{H}_{\rm isotr}^{\rm 6PN, D-B}$ is directly related to 
those in Eqs.~(\ref{eq:CB}, \ref{eq:CD}).
We now try to construct a canonical transformation from our 6PN Hamiltonian in harmonic coordinates to $O(G_N^3)$
to both the 6PN expanded Hamiltonians (\ref{eq:H6PNexpF}) implied by either $f_6^{\rm B}(\nu)$ or $f_6^{\rm D}(\nu)$.

If using $f_6^{\rm B}$, the generator
%-----------------------------------------------------------------------------------------------------------------------
\begin{eqnarray}
%-------------------------------------
g_{6} &=& \frac{p.r}{r} \Biggl\{
\Biggl(
        \frac{59 \nu ^3}{32}-\frac{4101 \nu ^4}{512}+\frac{2133 \nu ^5}{256}-\frac{21 \nu ^6}{1024}\Biggr) p^{10}
+\Biggl(
        -\frac{221 \nu ^3}{512}+\frac{115 \nu ^4}{64}-\frac{1091 \nu ^5}{768}
\nonumber\\ &&
-\frac{7 \nu ^6}{1024}\Biggr) p^8 (p.n)^2
+\Biggl(
        \frac{177 \nu ^3}{2560}-\frac{111 \nu ^4}{640}-\frac{1319 \nu ^5}{5120}-\frac{21 \nu ^6}{5120}\big) p^6 
(p.n)^4
+\Biggl(
        \frac{11 \nu ^3}{512}-\frac{23 \nu ^4}{128}
\nonumber\\ &&
+\frac{463 \nu ^5}{1024}-\frac{3 \nu ^6}{1024}\Biggr) p^4 (p.n)^6
+\Biggl(
        -\frac{7 \nu ^3}{512}+\frac{49 \nu ^4}{512}-\frac{49 \nu ^5}{256}-\frac{7 \nu ^6}{3072}\big) p^2 (p.n)^8
+\frac{21 \nu ^6}{1024} (p.n)^{10}
\nonumber\\ &&
%---
+\frac{1}{r} \Biggl[
\Biggl(
                \frac{14557 \nu }{512}-\frac{90189 \nu ^2}{512}+\frac{21797 \nu ^3}{64}-\frac{97743 \nu ^4}{512}-\frac{2917 
\nu ^5}{256}+\frac{7 \nu ^6}{512}\Biggr) p^8
+ \Biggl(
                -\frac{1943 \nu }{384}
\nonumber\\ &&
+\frac{49709 \nu ^2}{768}-\frac{71357 \nu ^3}{384}+\frac{233293 \nu 
^4}{1536}+\frac{83461 \nu ^5}{1536}+\frac{\nu ^6}{192}\Biggr) p^6 (p.n)^2
+        \Biggl(
                -\frac{5021 \nu }{1920}-\frac{27691 \nu ^2}{3840}
\nonumber\\ &&
+\frac{9845 \nu ^3}{256}-\frac{127735 \nu 
^4}{1536}-\frac{140303 \nu ^5}{7680}+\frac{19 \nu ^6}{3840}\Biggr) p^4 (p.n)^4
        + \Biggl(
                \frac{7783 \nu }{8960}+
                \frac{11253 \nu ^2}{896}-\frac{75627 \nu ^3}{2240}
\nonumber\\ &&
+\frac{1258391 \nu ^4}{17920}+\frac{181733 \nu 
^5}{17920}+\frac{3 \nu ^6}{320}\Biggr) p^2 (p.n)^6
        + \Biggl(
                -\frac{5333 \nu }{11520}-\frac{40343 \nu ^2}{11520}+\frac{226997 \nu ^3}{20160}
\nonumber\\ &&
-\frac{843847 \nu 
^4}{32256}-\frac{76493 \nu ^5}{161280}+\frac{77 \nu ^6}{1536}\Biggr) (p.n)^8
\Biggr]
%---
+\frac{1}{r^2} \Biggl[
         \Biggl(
                \frac{2207017 \nu }{17640}
                -\frac{11876924429 \nu ^2}{50803200}
\nonumber\\ &&         
       -\frac{3077092201 \nu ^3}{2903040}
                +\frac{1347718537 \nu ^4}{2822400}
                -\frac{53671 \nu ^5}{1024}
                +\frac{7 \nu ^6}{128}
                +\Biggl(\frac{161901 \nu }{131072}
                -\frac{32955 \nu ^2 }{8192}
                +\frac{7023 \nu ^3 }{512}
\nonumber\\ &&                
-\frac{213837 \nu ^4 }{8192} \Biggr) \pi^2
                + \Biggl(
                        \frac{17331 \nu }{70}-\frac{780211 \nu ^2}{720}+\frac{1631395 \nu ^3}{1008}-\frac{2975597 \nu 
^4}{5040}\Biggr) L
        \Biggr) p^6
\nonumber\\ &&
+        \Biggl(
                -\frac{128052787 \nu }{1693440}
                -\frac{263026853 \nu ^2}{211680}
                +\frac{1871276137 \nu ^3}{1354752}
                -\frac{13422417097 \nu ^4}{1693440}
                -\frac{8877829 \nu ^5}{46080}
\nonumber\\ &&
                +\frac{181 \nu ^6}{11520}
                +\Biggl(\frac{621015 \nu}{131072}
                -\frac{12495 \nu ^2}{2048}
                +\frac{277755 \nu ^3}{8192}
                -\frac{1405305 \nu ^4}{8192} \Biggr) \pi^2
                + \Biggl(
                        \frac{639 \nu }{7}+\frac{6851 \nu ^2}{14}
\nonumber\\ &&
+\frac{7773 \nu ^3}{112}-\frac{38511 \nu ^4}{112}\Biggr) L
        \Biggr) p^2 (p.n)^4
        + \Biggl(
                -\frac{159648631 \nu }{4233600}
                +\frac{1978082033 \nu ^2}{1209600}
                -\frac{55736243399 \nu ^3}{33868800}
\nonumber\\ &&
                +\frac{66817696627 \nu ^4}{16934400}
                +\frac{1544101 \nu ^5}{9216}
                +\frac{23 \nu ^6}{2304}
                +\Biggl(-\frac{398991 \nu}{131072}
                +\frac{33957 \nu ^2}{8192}
                -\frac{356457 \nu ^3}{8192}
\nonumber\\ &&           
     +\frac{525735 \nu ^4}{4096}\Biggr) \pi^2
                 +\Biggl(
                        -\frac{5403 \nu }{140}-\frac{5359 \nu ^2}{35}-\frac{17339 \nu ^3}{40}
                +\frac{69603 \nu ^4}{560}\Biggr) L
       \Biggr) p^4 (p.n)^2
\nonumber\\ &&        
+ \Biggl(
                -\frac{4066361 \nu }{120960}
                +\frac{53428687 \nu ^2}{103680}
                -\frac{233917561 \nu ^3}{967680}
                +\frac{1274653087 \nu ^4}{290304}
                +\frac{4666447 \nu ^5}{64512}
                +\frac{469 \nu ^6}{11520}
\nonumber\\ &&             
   +\Biggl(-\frac{325045 \nu}{131072}
                -\frac{2625 \nu ^2}{4096}
                +\frac{87185 \nu ^3}{4096}
                +\frac{1015 \nu ^4}{32}\Biggr) \pi^2
                +\Biggl(
                        -111 \nu -\frac{212 \nu ^2}{3}-\frac{580 \nu ^3}{3}
\nonumber\\ &&
+\frac{1789 \nu ^4}{3}\Biggr) L
        \Biggr) (p.n)^6
\big)
\Biggr\}
\end{eqnarray}
%--------------------------------------------------------------------------------------------------------
establishes a canonical transformation to 6PN.

If we use $f_6^{\rm D}$, we find no solution for the linear system  determining the parameters $\alpha_{k,l,j}$
in Eq.~(\ref{eq:gstr}). Investigating this closer, it turns out, that the contributions to the expanded Hamiltonian 
in isotropic coordinates $\propto p^8 u^3$ are already determined by the parameters $\alpha_{k,l,j}$ fixing the remaining
part of the transformation from harmonic to isotropic coordinates. This applies to the complete structure in $\nu$ at 
$O(p^8 u^3)$. The replacement (\ref{eq:f6D}) does then not allow to find a canonical transformation. 
%----------------------------------------------------------------------------------------------------------------
\section{Conclusions}
\label{sec:5}
%----------------------------------------------------------------------------------------------------------------

\vspace*{1mm}
\noindent
Within an effective field theory approach to the Einstein--Hilbert Lagrangian \cite{Kol:2007bc} we have calculated
the effective Hamiltonian at the sixth post--Newtonian order for all contributions to $O(G_N^3)$ ab initio for the 
first time. The overall computation time for this project has been one month.
For these terms we agree with all results in the literature to the level of 5PN. We also agree to the 
post--Newtonian expansion of the 3rd post--Minkowskian results of \cite{Bern:2019nnu,Bern:2019crd}, but do not
confirm the 6PN contribution of $O(\nu^2 p^8 u^3)$ of Ref.~\cite{Damour:2019lcq} obtained for the scattering angle.
We have applied the method of canonical transformations to investigate the equivalence of different effective 
Hamiltonians. Whenever these transformations exist one assures that all observables derived from the respective dynamics
are the same.

It is fully justified and necessary to critically investigate the different computation methods used to obtain 
theoretical predictions for the observables characterizing the process of the coalescence of two massive astrophysical 
objects within general relativity. All of these calculations, having now reached already an unprecedented level
of precision, are technically difficult and require cutting edge methods making a continuous monitoring necessary.
In this context, techniques having been developed in relativistic quantum field theory, turn out to be very useful
and have high relevance for loop calculations within effective field theories, as those applied here to a classical theory 
such as Einstein gravity.

\vspace*{5mm}
\noindent
{\bf Acknowledgment.} We thank Th.~Damour, J.~Steinhoff and J.~Vines for discussions. This work has 
been funded in part by EU TMR network SAGEX agreement No. 764850 (Marie Sk\l{}odowska-Curie) and COST action 
CA16201: Unraveling new physics at the LHC through the precision frontier. G.~Sch\"afer has been supported in 
part by Kolleg Mathematik Physik Berlin (KMPB). Part of the text has been typesetted using {\tt SigmaToTeX} of the 
package {\tt Sigma} \cite{SIG1,SIG2}.\footnote{{\sf Note added.} After the completion of this paper we have discussed our result with Th.~Damour, who reported to us
that a paper of his is in preparation \cite{THD} which also comes to the conclusion that the results of 
\cite{Bern:2019nnu} hold to 6PN and $O(G_N^3)$.}

\vspace*{5mm}
\noindent

%-------------------------------------------------------------------------------------

%-------------------------------------------------------------------------------------
\end{document}